\pdfoutput=1
\documentclass[reprint,twocolumn, linenumbers,aip,graphicx,amsmath]{revtex4-1}
\usepackage{graphicx}
\usepackage{dcolumn}
\usepackage{bm}
\usepackage[pdftex]{hyperref}
\usepackage[pdftex]{hyperref}
\usepackage{hyperref}
\hypersetup{
    colorlinks,%
    citecolor=blue,%
    filecolor=black,%
    linkcolor=black,%
    urlcolor=blue
}
\usepackage{url}
\draft
\begin{document}


\title{ Dielectric environment mediated  quantum screening of one dimensional electron gas} 



\author{Aniruddha Konar}
\email[]{akonar@nd.edu}
\affiliation{ Department of Physics, University of Notre Dame, Indiana 46556, USA.}
\affiliation{ Midwest Institute for Nanoelectronics Discovery (MIND), Indiana 46556, USA.}
\author{ Tian Fang}
\affiliation{ Department of Electrical Engineering, University of Notre Dame, Indiana 46556, USA.}
\affiliation{ Midwest Institute for Nanoelectronics Discovery (MIND), Indiana 46556, USA.}
\author{ Debdeep Jena}
\affiliation{ Department of Electrical Engineering, University of Notre Dame, Indiana 46556, USA.}
\affiliation{ Midwest Institute for Nanoelectronics Discovery (MIND), Indiana 46556, USA.}


\date{\today}

\begin{abstract}
 Relaxing the assumption of `` infinite and homogenous background''  the dielectric response function  of one-dimensional (1D) semiconducting  nanowires embedded in a  dielectric environment is calculated. It is shown that high-$\kappa$ (higher than semiconductor dielectric constant) dielectric environment reduces the screening by the free carriers inside the nanostructure whereas, low dielectric environment increases the Coulombic interaction between free carriers and enhances the strength of screening function. In long wavelength limit, dielectric screening and collective excitation of electron gas are found to be solely determined by the environment instead of the semiconductor. Behavior of static dielectric function is particularly addressed at a specific wavevector $q=2k_{F}$;  a wavevector ubiquitously appears in charge transport in nanostructures. 
 \end{abstract}

\maketitle 

Low-dimensional structures such as semiconducting nanowires (1D) are being investigated intensively for their potential applications in high-speed electronic and optical devices \cite{LiReview06}.  These semiconducting nanowires either can be freestanding or can be coated with different dielectric environment appropriate to device application. For example, in nanowire-based field effect transistors (FET), wires are usually coated with high-$\kappa$ dielectrics ( HfO$_{2}$, ZrO$_{2}$ etc.) \cite{RoddaroAPL08} for improved charge control as well as for high electron mobility \cite{KonarJAP07}. On the other hand, for exciton-based devices, use of low-$\kappa$ (lower than semiconductor dielectric constant $\epsilon_{s}$) dielectric is beneficial as it enhances excitonic binding energy \cite{KeldyshJETP79}. These advantages in electronic and optical properties stems out from the fact that the Coulombic interaction between carrier and/or impurities inside the nanowires can be altered by altering the environment.  This tunability of the carrier-carrier interaction by dielectric environment is expected to modify many body effects such as dielectric screening by one dimensional electron gas (1DEG) inside the nanowire.\\
\par
Dielectric screening by free carriers plays a crucial role in determination of transport quantities (conductivity, mobility, etc) of a nanostructure. In a scattering event, the momentum-relaxation time ($\tau$) strongly ($\tau\sim\left|\epsilon(q,0)\right|^{-2}$) depends on the free electron screening inside the semiconductor. Hence an accurate knowledge of dielectric screening is necessary for a precise prediction of transport coefficients of a nanowire. The dielectric function of a semiconductor nanowire is composed of i) ionic ($\epsilon^{ion}$) and ii) electronic ($\epsilon^{ele}$) contributions. $\epsilon^{ion}$ is a inherent property (crystal property) of semiconductors, while $\epsilon^{el}$  (commonly known as screening function) depends on the magnitude of the electron-electron interaction inside a material. As dielectric environment can alter the Coulomb potential inside a nanowire, it is expected that dielectric environment will have a pronounce effect of the free electron screening of the nanowire\cite{note}. Previous models \cite{WilliamsPRB74,LeeJAP84,LiPRB89,LiPRB91} for dielectric function of 1DEG assumes that the electron gas has a infinite homogenous background having dielectric constant ($\epsilon_{s}$) same as the semiconductor. For a nanowire of few nm radius, ``infinite background'' approximation certainly breaks down and at the nanowire/environment interface ``homogenous background'' assumption fails. In this work, assumption ``infinite homogenous background'' is relaxed, and incorporating dielectric mismatch factor at the nanowire/environment a consistent theory of dielectric function is developed using the method of ``{\it{self consistent field''}}\cite{CohenPR59} (also known as random-phase approximation or RPA).
\par
We consider an infinitely long semiconductor wire (dielectric constant $\epsilon_{s}$) of a radius ($R$) few nanometers embedded in a dielectric (dielectric constant $\epsilon_{e}$) environment.  To investigate the dielectric response of the electron gas inside the wire, we place an oscillating test charge at ({\bf{r}}$_{0}$,z$_{0}$)=(0,0) of density $n_{0}(r,t)=e\delta({\bf{r}})e^{-i\omega t}$. This test charge creates a longitudinal electric field $V_{0}(r,z)e^{-i\omega t}$ in the nanowire and in response to this perturbation, free electrons inside the nanowire rearrange themselves to screen the field. The resultant Hamiltonian of electrons confined in the wire is $H=H_{0}+ V({\bf{r}},t)$, where $V({\bf{r}},t)$ is the self-consistent potential in response to the external perturbation $V_{0}({\bf{r}},t)$.  The unperturbed single-particle Hamiltonian $H_{0}={\bf{p}}^{2}/2m^{\star} +V_{con}(r)$ satisfy Schroedinger equation $H_{0}|n,k\rangle={\mathcal {E}}_{n,k}|n,k\rangle$. Here $m^{\star}$ is the effective mass of electron, $k$ is the one dimensional wave vector, $|n,k\rangle$ and ${\mathcal{E}}_{n,k}$ are the eigenvectors and eigen-energy of the unperturbed Hamiltonian and $V_{con}(r)$ is the confinement potential for electrons inside the nanowire.  Assuming electrons are confined in a infinite-barrier potential, the eigen-energies are ${\mathcal{E}}_{n,k}={\mathcal{E}}_{n}+\hbar^{2}k^{2}/2m^{\star}$, where ${\mathcal{E}}_{n}$ is the ground state energy of the $n$th 1D subband and $\hbar$ is the reduced Planck constant. The corresponding wavefunction is $\Psi_{n,k}(r,z)=\langle r|n,k\rangle=\phi_{n}(r)\cdot[\exp(ikz)/\sqrt{L}]$, where $\phi_{n}(r)$ is the radial part and $L$ is the length of the nanowire. The dielectric function of an electron gas is defined by the relation\cite{BookFerry}
\begin{equation}
V_{nn'}=\sum_{mm'}\epsilon^{-1}_{nn',mm'}(q,\omega)V_{mm'}^{0},
\end{equation}
where $\epsilon^{-1}_{nn',mm'}(q,\omega)$ is the four dimensional dielectric matrix  and $V_{ij}(V^{0}_{ij})=\langle j,k+q|V(V_{0})|i,k\rangle$ are the transition matrix element between states $|i,k\rangle$ and $|j,k+q\rangle$. Diagonal elements of the dielectric matrix represent the intrasubbabd polarization of the 1DEG whereas, the off-diagonal terms coming from inter-subband transitions. In size quantum limit (SQL) of nanowire, when carriers are confined in the lowest ground state and intersubband separation is large, dielectric function becomes a scalar quantity.

\par
The self-consistent potential contains both original perturbation as well as the screened potential by the mobile charges, i.e. $V({\bf{r}},t)=V_{0}({\bf{r}},t)+V_{s}({\bf{r}},t)$. For the evaluation of the  dielectric response of a 1D electron gas, it is imperative to calculate the screening potential $V_{s}$ (see eq. 1)). The self-consistent potential $V({\bf{r}},t)$, upon acting on state $|n,k\rangle$ mixes in other state such that wave function becomes $\Psi(r,t)=|n,k\rangle+b_{k+q}(t)|n',k+q\rangle$. The coefficient $b_{k,k+q}(t)$ is given by time dependent perturbation theory\cite{BookZiman}
\begin{equation}
b_{k,k+q}(t)=\frac{V_{nn'}(q)e^{-i\omega t}}{{\mathcal{E}}_{n'}(k+q)-{\mathcal{E}}_{n}(k)-\hbar\omega},
\end{equation}
where, $V_{nn'}=\langle n',k+q|V|n,k\rangle$ is the matrix element between state $|n,k\rangle$ and $|n',k+q\rangle$. The perturbation-induced charge density is $n^{ind}(r,t,z)=-2e\sum_{k,nn'}f^{0}_{n}(k)\Big[|\Psi(r,t)|^{2}-|\Psi_{n,k}(r,z)|^{2}\Big]$, where, $e$ is the charge of an electron and $f^{0}_{n}(k)$ denotes the equilibrium Fermi-Dirac occupation probability of a state $|n,k\rangle$ such that $2\sum_{n,k}f^{0}_{n}(k)=n_{1d}$, $n_{1d}$ being the equilibrium homogeneous unperturbed electron gas density. Assuming that the perturbation is weak enough such that response is linear and neglecting terms $b^{2}_{n,k+q}$ and higher orders,  induced charge density can be written as $n^{ind}(r,t)=--e\sum_{nn'}\phi_{n}(r)\phi_{n'}(r)V_{nn'}{\mathcal{F}}_{nn'}(q,\omega)e^{iqz}e^{i\omega t}$, where ${\mathcal{F}}_{nn}(q,\omega)$ is the polarization function \cite{BookZiman} (Lindhard function) obtained by summing the Feynman diagram of electron-electron interaction containing single fermion loop \cite{WilliamsPRB74,BookMahan},
\begin{equation}
{\mathcal{F}}_{nn'}(q,\omega)=\frac{2}{L}\sum_{k}\frac{f^{0}_{n}(k)-f^{0}_{n'}(k+q)}{{\mathcal{E}}_{n'}(k+q)-{\mathcal{E}}_{n}(k)-\hbar\omega}.
\end{equation}
Note that, the induced charge density has the same harmonic dependence as the self consistent potential. The induced charge density is related to the screening potential by  Poisson's equation $\nabla^{2}V_{s}({\bf{r}})=-n^{ind}({\bf{r}})/\epsilon_{0}\epsilon_{s}$, where $\epsilon_{0}$ is the free-space permittivity. Expressing screening potential in Fourier components $V_{s}(r,z)=\sum_{-\infty}^{\infty}v_{s}(r,q)e^{iqz}$, where $q=k'-k$, one obtain the differential equation for the screening potential
\begin{eqnarray}
\frac{1}{r}\frac{d}{dr}\Big(r\frac{dv_{s}}{dr}\Big)-q^{2}v_{s}=\begin{cases}
-n^{ind}(r)/\epsilon_{0}\epsilon_{s}, &  r \leq R\\
0,  & r \ge R.
\end{cases}
\end{eqnarray}
The Green's function appropriate to above differential equation with dielectric mismatch effect is \cite{MuljarovPRB00,KonarJAP07}
\begin{eqnarray}
G(r,r',q)&=&
\frac{1}{\pi}\Big[\underbrace{I_{0}(q.r_{<})K_{0}(qr_{>})}_{g^{inhom}(r,r')}+\underbrace{{\mathcal{U}}(qR)I_{0}(qr)K_{0}(qr')}_{g^{hom}(r,r')}\Big]\nonumber\\
{\mathcal{U}}(x)&=&\frac{(\epsilon_{s}-\epsilon_{e})K_{0}(x)K_{1}(x)}{\epsilon_{e}I_{0}(x)K_{1}(x)+\epsilon_{s}I_{1}(x)K_{0}(x)}
\end{eqnarray}
where, $g^{hom(inhom)}(r,r')$ is the homogenous (inhomogenous) part of the Green's function, $r_{<(>)}=$min(max)$[r,r']$, $I_{n}(..)$ and $K_{n}(...)$ are the nth order modified Bessel functions. For large $x$ $ (x>\sqrt{n^{2}-1})$, $I_{n}(x)\approx e^{x}/\sqrt{2\pi x}$, $K_{n}(x)\approx e^{x}\sqrt{2\pi/ x}$ and the function ${\mathcal{U}}(qR)\to ( \pi\gamma/2)e^{-2qR}$, where $\gamma=(\epsilon_{s}-\epsilon_{e})/(\epsilon_{s}+\epsilon_{e})$ is the dielectric mismatch factor. The tunability of the strength of the Green's function comes through its dependence on $\gamma$, which enhances (reduces) the strength for $\epsilon_{s}>\epsilon_{e} (\epsilon_{s}<\epsilon_{e})$. For an infinite homogeneous environment ($\epsilon_{e}=\epsilon_{s}$), $\gamma=0$, and the Green's function is independent of the dielectric environment.  Using the above Green's function, the induced potential inside the nanowire can be written as $v_{s}(r,q)=e^{2}/4\pi\epsilon_{0}\epsilon_{s}\int_{ 0}^{R}G(r,r',q)n^{ind}(r')r'dr$\cite{BookJackson}. In the size quantum limit (SQL), the nanowire is thin,  ($R< \lambda_{dB}, \lambda_{dB}$ is de Broglie wavelength of an electron) and only the lowest subband is populated. Moreover, for a thin nanowire, inter-subband separation energy is large ($\Delta{\mathcal{E}}_{n}\propto 1/R^{2}$) such that inter-subband transition can be neglected ($n=n'=1$).  In such a scenario (SQL limit), the dielectric matrix becomes scalar, i.e. $\epsilon_{nn'}(q,\omega)\rightarrow\epsilon_{11}(q,\omega)$.  Assuming $\phi_{n=1}(r)\approx 1/\sqrt{\pi R^{2}}$, the dynamic dielectric function of an 1DEG at temperature $T=0$ is\cite{relation}
\begin{eqnarray}
\epsilon_{1d}(q,\omega,{\mathcal{E}}_{F})&=&1-\frac{e^{2}}{4\pi\epsilon_{0}\epsilon_{s}V_{11}}\int_{0}^{R}\phi_{1}^{2}(r)r\int_{0}^{R}G(r,r')n^{ind}(r')r'dr'dr\nonumber\\
&=&1+\frac{1}{\pi a^{\star}_{B}R^{2}}\frac{F(x)}{q^3}ln\left|\frac{(q+2k_{F})^2-(\frac{2m^{\star}\omega}{\hbar q})^2}{(q-2k_{F})^2-(\frac{2m^{\star}\omega}{\hbar q})^2}\right|
\label{eq6}
\end{eqnarray}
where $x=qR$ a dimensionless quantity, $F(x)= \Big[\frac{1}{2}+I_{1}(x)[{\mathcal{U}}(x)I_{1}(x)-K_{1}(x)]\Big]$, $a_{B}^{\star}=4\pi\epsilon_{0}\epsilon_{s}/m^{\star}e^{2}$ is the effective bulk Bohr radius, $k_{F}=\pi n_{1d}/2$ is the Fermi wavevector and ${\mathcal{E}}_{F}=\hbar^{2}k_{F}^{2}/(2m^{\star})$ is the corresponding Fermi energy.  The logarithmic term in Eq.\ref{eq6} is arising from the Lindhard function ${\mathcal{F}}_{11}(q,\omega)$ which has been evaluated analytically in SQL \cite{WilliamsPRB74}.
\begin{figure}[t]%
\includegraphics*[width=85 mm]{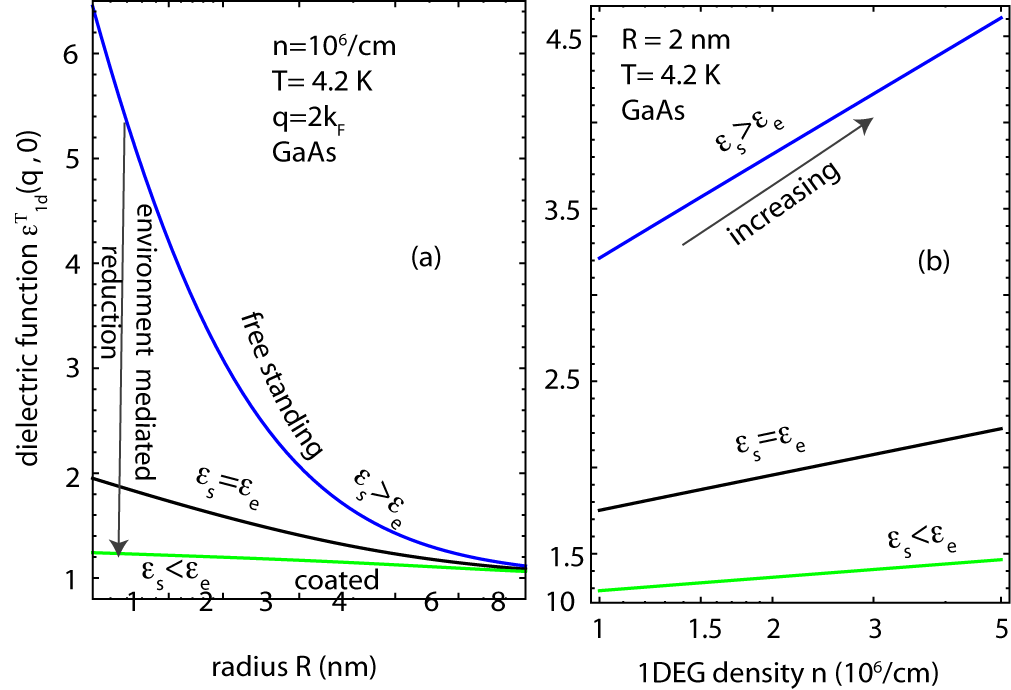}
\caption{%
  dielectric function of a nanowire a) with nanowire radius ($R$) and b) as a function of carrier density ($n$) for three different dielectric environments of $\epsilon_{e}=1$ (upper branch), $\epsilon_{e}=\epsilon_{s}=13$ (middle) and $\epsilon_{e}=100$ (lower branch). }
\label{Fig1}
\end{figure}
In the context of charge transport inside the nanowire, static dielectric function $\epsilon_{1d}(q,\omega=0)$ is relevant than dynamic one. In the long wavelength ($q\ll 2k_{F}$) limit,  static dielectric function $\epsilon_{1d}(q,0)$ for a thin nanowire ($qR\rightarrow 0$) is
 \begin{equation}
\epsilon_{1d}(q,0)=1-\frac{e^{2}}{2\pi \epsilon_{0}{\bf{\epsilon_{e}}}}\left[\ln\left(qR\right)\right]{\mathcal{D}}_{1d}({\mathcal{E}}_{F}),
 \end{equation}
where, ${\mathcal{D}}_{1d}({\mathcal{E}}_{F})=1/\pi\hbar\sqrt{2m^{\star}/{\mathcal{E}}_{F}}$ is thw 1D density of states per unite length at Fermi energy ${\mathcal{E}}_{F}$. In sharp contrast to previous models\cite{BookFerry}, $\epsilon_{e}$ instead of $\epsilon_{s}$, determines the long-wavelength behavior of the static dielectric function. 

For large momentum $(q>>2k_{F}$), $\epsilon_{1d}(q,0)\to 1$ as the second term of Eq. 6 falls off rapidly ($q^{-5}$) with q. For a degenerate 1DEG in SQL , only possible way of scattering is backscattering which leads to a momentum transfer $q=2k_{F}$ in any intrasubband elastic scattering process. As a result, $\epsilon_{1d}(q=2k_{F},0)$ plays an important role in momentum relaxation rate calculation. 
In the static limit  ($\omega=0)$,  the dielectric function $\epsilon_{1d}(q,0)$ at $T=0$ is singular for $q=2k_{F}$. This divergence is related to Pierl's instability which is a characteristic signature of a 1DEG. At finite temperature, smearing of Fermi function removes this singularity.  At finite temperature, the static dielectric function is given by Maldague's prescription\cite{MaldagueSS78}
\begin{equation}
\epsilon_{1d}^{T}(q,0)=\int_{0}^{\infty}d{\mathcal{E}}\epsilon_{1d}(q,0,{\mathcal{E}})\left[4k_{B}T\cosh^{2}\left[\frac{{\mathcal{E}}-{\mathcal{E}}_{F}}{2k_{B}T}\right]\right]
\end{equation}\\
\par
\begin{figure}[t]%
\includegraphics*[width=90 mm]{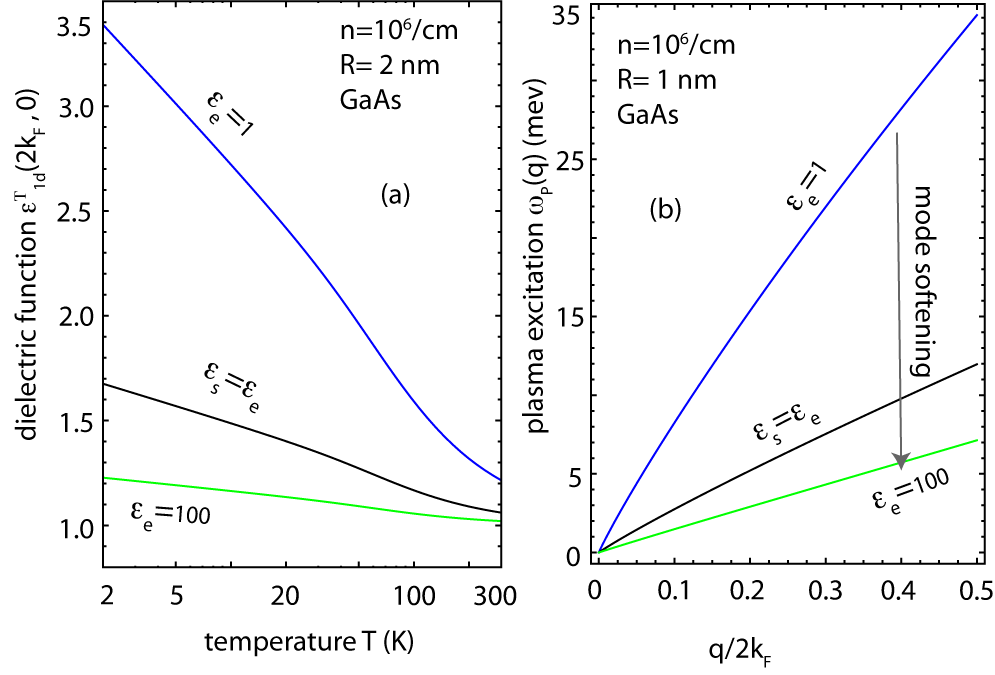}
\caption{%
 a) dielectric function of a nanowire  with temperature ($T$) and b) plasma frequency of an 1DEG with wavevector ($q$) for three different dielectric environments. }
\label{Fig2}
\end{figure}

Fig.\ref{Fig1}a) shows the static dielectric function of a GaAs nanowire at $q=2k_{F}$ with nanowire radius $R$ for three different dielectric medium. Note that even negligible smearing of Fermi distribution at $T=4.2$ K is enough to remove the divergence at $q=2k_{F}$.  For  coated nanowires with $\epsilon_{e}>\epsilon_{s}$, dielectric screening is strongly reduced as shown in Fig. 2 b). At large radius ($R>>1/4k_{F}$), nanowire tends to the bulk structure and the dielectric mismatch effect on the screening function vanishes. With increasing carrier density, dielectric screening inside the nanowire increases (see Fig. 1b) maintaing the effect of dielectric environment intact. At higher carrier densities, more than one subband is populated and inter-subband contribution to the total dielectric function should be taken account for a complete description of free electron screening inside the nanowire. With increasing temperature, thermal fluctuation reduces the free electron screening inside the nanowire and the effect of environmental dielectric on the screening function is partially washed away (see Fig. \ref{Fig2}a).
  \par
 As the dynamic ($\omega \neq 0$) dielectric function $\epsilon_{1d}(q,\omega)$ contains the dielectric mismatch factor, collective excitation of 1DEG is also expected to depend on the dielectric environment. Collective excitation of a electron gas is defined the pole of the full dynamic dielectric function, i.e. by $\epsilon_{1d}(q,\omega_{p})=0$, where $\omega_{p}$ is the plasma frequency of the electron gas.  Fig.\ref{Fig2}b) shows the plasma dispersion of intra-subband collective excitation of a thin nanowire ($R=2$nm) for different dielectric environments. For $q<1/2R$, dielectric environment has finite effect on the collective excitation frequency of 1DEG. The softening of plasma frequency with high-$\epsilon_{e}$ dielectric environment is the consequence of the reduction of Coulomb interaction between electron and positive background, which acts as a restoration force of the collective oscillation of the electron gas.  For small $q$, frequency of collective excitation goes to zero for all dielectric environment following the relation $\omega_{p}(q)\approx \omega_{0}q\sqrt{ -\ln(qR)}$, where $\omega_{0}=\sqrt{n_{1d}e^{2}/(4\pi\epsilon_{o}{\bf{\epsilon_{e}}}m^{\star})}$.  Note the explicit appearance of $\epsilon_{e}$ in $\omega_{0}$ justifies the role of environment in collective excitation of 1DEG inside the wire.  \\
 \par
 The length scale at which dielectric environment plays an important role can be determined by investigating the behavior of ${\mathcal{U}}(qR)$. For large $qR$, ${\mathcal{U}}(qR)\sim e^{-4k_{F}R}$. Hence for $R>>1/(4k_{F})$, ${\mathcal{U}}(qR)$ becomes negligible and dielectric effect vanishes. For a numerical estimates, at carrier density $n_{1d}=10^{6}$ /cm dielectric effect vanishes for $R>> 2$ nm, whereas at lower density ($n_{1d}=10^{5}$ /cm) environmental effect on quantum screening function persists for wire radius $R\approx20$ nm.\\
 \par
 In our work we assume an infinite confining potential for electron inside the wire. Relaxing this assumption will results in electron mass enhancement due to leaking of wavefunction into the barrier. For high-$\kappa$ oxides the typical barrier height is $\sim 1$ eV, for which nominal increase in electron mass can be neglected \cite{JenaPRL07}. The assumption of constant radial part of the wavefunction is justified for thin nanowires. Choosing a different form for the radial part will change the absolute value of screening function for thick (for large $R$ dielectric environment effect ceases out anyway) wires keeping the relative effect of environments unchanged. \\
  In conclusion, we have shown that the free electron screening inside a nanowire depends on the environment surrounding it. For a nanowire coated with high-$\kappa$ dielectric, perturbation inside the nanowire is poorly screened compared to a freestanding nanowire. It is shown that  both static dielectric function and plasma dispersion at long-wavelength limit gets modified by environment . We derived the length-scale at which the environment has meaningful effect on the electron gas inside the nanowire.  Results are analytical and will be useful for accurate prediction of transport coefficients in nanowire-based electronic devices. \\
 \par
 The authors would like to acknowledge  National Science Foundation (NSF) NSF Grant Nos. DMR-0907583 and NSF DMR-0645698), Midwest Institute for Nanoelectronics Discovery (MIND) for the financial support for this work.  


\begin{thebibliography}{99}
\bibitem{note} One such example is  dielectric environment mediated free electron screening in graphehe, where an average dielectric constant of graphene $\epsilon_{gr}^{avg}=(\epsilon_{e}^{t}+\epsilon_{e}^{b})/2 $ is used in Thomas-Fermi screening function. Here $\epsilon_{e}^{t} (\epsilon_{e}^{b})$ is the dielectric constant of top (bottom) environment of graphene and the crystal property of atomically thin (monolayer 0.3 nm) graphene is neglected. For a nanowire of radius few nm, crystal property of the semiconductor can not be neglected and a rigorous model containing both $\epsilon_{e}$ and $\epsilon_{s}$ is necessary.
\bibitem{relation} For evaluation of the integral the identity $I_{n}(x)K_{n+1}(x)+ I_{n+1}(x)K_{n}(x)=1/x $ is used.
\end{thebibliography}
\end{document}